\documentclass[11pt,a4paper]{article}
\pdfoutput=1

\usepackage{jheppub}
\usepackage[english]{babel}
\usepackage[utf8]{inputenc}
\usepackage{epsf}
\usepackage{graphicx}

\usepackage{verbatim} % allows block comments

\usepackage{amsmath}
\usepackage{amssymb}
\usepackage{mathrsfs}

\usepackage{mathbbol}
\usepackage{appendix}

\usepackage{amsfonts}
\usepackage{bbding}
\usepackage{array}
\usepackage{todonotes}
\usepackage{subfig}
\usepackage{tikz}

\DeclareRobustCommand{\swatch}[1]{\tikz[baseline=-0.6ex]\node[fill=#1,shape=rectangle,draw=black,thick,minimum width=5mm,rounded corners=0.5pt](){};}
\newcommand{\MET}{\ensuremath{E_T^{\rm miss}}}

\definecolor{black}{HTML}{000000}
\definecolor{blue}{HTML}{0000FF}
\definecolor{darkgoldenrod}{HTML}{B8860B}
\definecolor{magenta}{HTML}{FF00FF}
\definecolor{orange}{HTML}{FFA500}
\definecolor{saddlebrown}{HTML}{8B4513}
\definecolor{silver}{HTML}{C0C0C0}
\definecolor{skyblue}{HTML}{87CEEB}
\definecolor{turquoise}{HTML}{40E0D0}
\definecolor{yellow}{HTML}{FFFF00}

\usepackage{color} % colored text
\usepackage{tikz}
\usetikzlibrary{shapes,arrows,positioning,automata,backgrounds,calc,er,patterns}
\usepackage[compat=1.1.0]{tikz-feynman}
\vspace*{1cm}
\allowdisplaybreaks
\title{Probing exotic long-lived particles from the prompt side using the CONTUR method}

\author[1]{L. Corpe}
\author[1]{A. Goudelis}
\author[1]{S. Jeannot}
\author[2]{S. Jeon}

\affiliation[1]{Université Clermont Auvergne, CNRS/IN2P3, LPCA, 63000 Clermont-Ferrand, France}
\affiliation[2]{Boston University, 590 Commonwealth Avenue, Boston, Massachusetts 02215, USA}
\emailAdd{lcorpe@cern.ch}

\abstract{
A method to derive constraints on new physics models featuring exotic long-lived particles using detector-corrected measurements of prompt states is presented. The \textsc{Contur} workflow is modified to either account for the fraction of long-lived particles which decay early enough to be reconstructed as prompt, or to be sensitive to the recoil of such particles against a prompt system. This makes it possible to determine how many of  signal events would be selected in the \textsc{Rivet} routines which encapsulate the fiducial regions of dozens of measurements of Standard Model processes by the ATLAS and CMS collaborations. New constraints are set on several popular exotic long-lived particle models in the very short-lifetime or very long-lifetime regimes, which are often poorly covered by direct searches. The probed models include feebly-interacting dark matter, hidden sector models mediated by a heavy neutral scalar, dark photon models and a model featuring photo-phobic axion-like particles.
}
\begin{document}
\maketitle

%%%%%%%%%%%%%%%%%%%%%%%%%%%%%%%%%%%%%%%%%%%%%%%%%%%%%%%%%%%%%%%%%%%%%%%%%%%%%%%%%%%
%%%%%%%%%%%%%%%%%%%%%%%%%%%%%%%%%%%%%%%%%%%%%%%%%%%%%%%%%%%%%%%%%%%%%%%%%%%%%%%%%%%
%%%%%%%%%%%%%%%%%%%%%%%%%%%%%%%%%%%%%%%%%%%%%%%%%%%%%%%%%%%%%%%%%%%%%%%%%%%%%%%%%%%
\section{Introduction}\label{sec:intro}

The discovery of the Higgs boson in 2012 closed an important chapter of the history of particle physics. But despite all the triumph of the discovery, the Standard Model (SM) is not the end of the story of fundamental physics. Indeed, the lack of a convincing candidate for Dark Matter (DM), the issue of the hierarchy problem, the matter-antimatter asymmetry of the universe, and the established fact of neutrino oscillations, all point towards the existence of new, undiscovered particles. 
%There is also good reason to believe~\cite{} \todo{@Andreas is there a good ref for this?} that such particles could naturally lie near the electroweak symmetry-breaking scale, and as such, ought to be discoverable at the LHC.
Nevertheless, now that the first decade of LHC data has been scrutinized, no long-standing signs of new particles or interactions have come to light. This prompts the question: ``are there assumptions in the LHC research programme that could be hampering a discovery?'' Indeed, one possible blind spot is the existence of semi-stable new particles, which could travel far enough into the detector to leave neither missing transverse momentum signatures, nor be reconstructed as standard objects. Since the standard ATLAS and CMS reconstruction algorithms assume that activity started at the interaction point, many of these new signals could have been thrown away as noise. Besides the experimental motivation, these exotic Long-Lived Particles (LLPs) also have good theoretical credentials. Long lifetimes occur when a particle's width becomes small, typically because of a small matrix element (for example, small couplings or heavy virtual particles mediating the decay) or phase space (for example, due to small mass differences between parent and daughter particles). These mechanisms occur frequently in the SM, and it should be expected that they would occur in a wide variety of new physics models that resolve the fundamental questions facing particle physics. Examples range from supersymmetry~\cite{ArkaniHamed:2012gw,Giudice:1998bp,Barbier:2004ez,Csaki:2015fea,Fan:2011yu,Fan:2012jf} to hidden sector (HS)~\cite{Strassler:2006im,Strassler:2006ri,Chan:2011aa} and axion-like particle models~\cite{Brivio:2017ije}.

Over the last decade, there has been an upsurge in interest in exotic LLP searches at the LHC~\cite{Alimena:2019zri}, encompassing an impressive variety of signatures from displaced vertices and jets to anomalous ionization (see Refs.~\cite{ATLAS:2023cjw,CMS-EXO-22-020-new,ATLAS:2024ocv, CMS:2024trg, ATLAS:2023zxo, ATLAS:2024qoo} for recent examples). These searches have had to re-invent the way that particles are reconstructed in multipurpose experiments like ATLAS or CMS, to bypass trigger constraints, re-purpose standard reconstruction algorithms and control unusual backgrounds (such as cosmic rays or beam-induced activity). These dedicated searches can be correspondingly complex and time-consuming. Nevertheless, much progress has been made in recent years, and constraints on the production cross-sections (times branching fraction) of LLP production against the lifetime are set. Summary plots that give an overview of the experimental situation, for example for HS models can be found in Ref.~\cite{ATL-PHYS-PUB-2022-007} for ATLAS and Ref.~\cite{CMSsummar} for CMS.

However, despite this progress, the low-lifetime (or equivalently, early-decay) regime remains ill-covered. Many models have no constraints set by the multipurpose experiments in the region of $c\tau$ (the speed of light times mean proper lifetime) that straddles prompt and displaced decays. This is ostensibly because of the difficulty in reconstructing secondary vertices and the presence of sometimes overwhelming SM backgrounds such as multijet production. It is also the case that direct searches are limited by the physical size of the detector, and hence sensitivity drops off at large lifetimes.
%In a sense, in these regions the question of whether the LLPs are really LLPs becomes relevant: at least some fraction of particles' actual decay positions would be small enough to look like prompt particles. 
In this paper, we argue that existing measurements of prompt processes at the LHC can be used to constrain LLP production, for two complementary reasons. First, whatever the mean proper lifetime of an exotic LLP, the decay distribution is a falling exponential, and therefore, some fraction of LLPs will behave like promptly-decaying particles. Second, if LLPs are produced with a prompt particle, then the presence of that LLP could already be inferred from the effect of its recoil on the prompt particle's transverse momentum spectrum (regardless of the decay mode or lifetime of the LLP). Hence, some LLP signatures might already have been visible in well-studied final states for which precision measurements exist at ATLAS and CMS, and can therefore already be excluded without a dedicated search. This would liberate person power to study other regions in more detail, making better use of ATLAS and CMS analyst resources.
%Further, we will see that additional constraints can be derived for specific models, where assumptions about the production mode of LLPs necessarily lead to predictions of non-SM prompt activity that could have been detected.
We will describe a method for estimating these constraints using existing open source software, and produce new limits on a variety of benchmark LLP models which are studied at the LHC, sometimes constraining the same models in some lifetime regions for the very first time.

The rest of this paper is organised as follows. Section~\ref{sec:contur} reviews the established \textsc{Contur} method to extract constraints on new-physics models from the existing bank of LHC measurements. Section~\ref{sec:contur_for_llps} describes how this method normally breaks down for LLP models, and describes a workflow to remedy this drawback. Sections~\ref{sec:fimps},~\ref{sec:HS},~\ref{sec:HZZd} and~\ref{sec:ALP} then illustrate the application of the modified method to several benchmark LLP models commonly probed by ATLAS and CMS, extracting new constraints which complement the direct search programme. The models are chosen to illustrate a breadth of experimental signatures: they include examples of single production and pair production; leptonic and hadronic decays; and scenarios where the LLP is electrically charged or neutral. 
Our choice of template models and parameter ranges is not guided by theoretical considerations. Instead we work with models which are already employed by LHC experimental collaborations to be able to establish comparisons with existing searches.
Finally, we present our conclusions in Section~\ref{sec:conclusions}.

\section{Re-interpretation of measurements with CONTUR}\label{sec:contur}

There exist broadly two  ways to extract constraints on new models using existing LHC results. The first is to embed the original benchmark model (on which constraints have been set by an experimental paper) within a wider theoretical framework (examples include \textsc{SModelS}~\cite{Altakach:2023tsd} or standalone analyses like Ref.~\cite{Haisch:2023rqs}). The second option is to determine the kinematics of new final states predicted by a model, and approximate the signal efficiency for this new model in a target analysis. This can be achieved using either efficiency maps provided by the experimental collaborations, or runnable code snippets which reproduce the signal region, if necessary with relevant smearing applied to approximate the effect of the detector. In the latter category, many open source tools provide libraries of re-interpretable analyses, usually exploiting searches. One tool however, makes use instead of the bank of detector-corrected measurements at the LHC, thus avoiding the need to approximate the detector response (often the most costly and approximate step). This method also benefits from the fact that measurement papers almost invariably provide experimentalist-produced definitions of the signal regions in the \textsc{Rivet}~\cite{Bierlich:2024vqo} library, which is by construction synchronised with the digitized measurement information on the \textsc{HEPData}~\cite{Maguire:2017ypu,HEPData} analysis preservation portal. This tool is called \textsc{Contur}~\cite{Butterworth:2016sqg,Buckley:2021neu}, and is able to provide rapid insights into constraints on new models (see Refs.~\cite{Butterworth:2024eyr,Altakach:2022nuo,Butterworth:2022dkt} for recent examples using models predicting new prompt particles).

The \textsc{Contur} workflow is detailed in Ref.~\cite{Buckley:2021neu} but the key points are summarised here. The starting point is a new physics model encapsulated in the standard Universal FeynRules Output (UFO) format~\cite{Degrande:2011ua}. The UFO can be passed to a Monte Carlo Event Generator (MCEG) (with suitable model parameter choices) to produce \textsc{HEPMC}~\cite{Verbytskyi:2020sus} files, which contain the particle-level history of the interaction for a given number of generated collision events, after hadronisation.
In the next step, the \textsc{HEPMC} events are analysed in \textsc{Rivet}, passing them through dozens of ``routines'' that encapsulate the definitions of the fiducial regions of as many measurements of SM processes. 
The result is a set of histograms in the \textsc{YODA}~\cite{Buckley:2023xqh} format which show where the signal would have shown up in the LHC measurements.
Since \textsc{Rivet} is automatically synchronised with \textsc{HEPData}, it is trivial to then compare the size of the new signal (stacked on SM predictions) to the measured cross-sections in each bin, within the experimental uncertainties.
The final step is to run a statistical analysis, grouping measurements into orthogonal pools (defined by final state, experiment and centre-of-mass energy) to avoid double counting, and produce a CLs~\cite{Read:2002hq} exclusion value.  \textsc{Contur} acts as a book-keeping and steering layer for this workflow, and provides the final statistical analysis.
The power of this method is that while a dedicated search for a new signal with ATLAS or CMS can take years, existing constraints can already be determined in a few days using \textsc{Contur}. Indeed, running \textsc{Contur} on a single set parameter point takes usually a few hours, and a grid of a few hundred parameter points can easily be done overnight on a standard batch computing farm.

The same method can also be used to extract estimates of the constraints on a given model after the end of the high-luminosity LHC (HL-LHC) lifetime in an automated way, as first demonstrated in Ref.~\cite{Butterworth:2024eyr}. Indeed, if we assume that 3000/fb are collected by the end of the HL-LHC programme, we can scale the uncertainties by the square root of the ratio of the increase in integrated luminosity for a given analysis and recalculate the CLs values.
Indeed, statistical uncertainties scale with the square root of integrated luminosity. The extrapolation relies on the assumption that experimental systematic uncertainties would also scale this way, which can be deemed a little optimistic. This optimistic assumption is counterbalanced by pessimistic assumptions that no improvements in theoretical uncertainties would take place in the intervening decades, that no new final states would be measured, and that no additional phase space would be explored. Assuming that the pessimistic and optimistic assumptions cancel out, we can take the HL-LHC extrapolations from \textsc{Contur} with a certain degree of confidence so long as these caveats are kept in mind.

\section{Adapting \textsc{Contur} to LLP topologies}\label{sec:contur_for_llps}

Until now, the \textsc{Contur}  method has not been usable for LLP signatures. Indeed, the definitions of the analysis signal regions encapsulated in \textsc{Rivet} always assume prompt behaviour (as they should, since the measured SM processes are prompt). Hence,  the detector corrections for those analyses would not account for jets or tracks that would be lost due to a displaced starting point. It could even be that data events containing displaced activity are thrown away by the measurements because they resemble detector noise. The consequence is an over-estimate of the selection efficiency for LLP signals, and hence unreliable CLs exclusions. For this reason, models (or regions of the model space) predicting LLPs could not be probed using the technique described in Section~\ref{sec:contur}. This prohibition is easier to state than to enforce however: it may not be immediately obvious to a user what combinations of parameters would lead to  particles with small widths and therefore long lifetimes. Indeed,  the width of a particle can depend on the interplay between the masses of daughter and mother particles, and the values of various couplings. When performing a scan over a range of parameters in a model with \textsc{Contur}, it is not trivial to determine which regions should be avoided, unless special care is taken, as was done for example in Ref.~\cite{Amrith:2018yfb}. Moreover, as we will show in what follows, some parameter space regions predicting long mean proper lifetimes should actually not be avoided, since meaningful constraints can be extracted by focusing on the fraction of LLPs decaying promptly (given the probabilistic nature of particle decay when seen on an event by event basis). A workflow to automatically detect LLPs in new models and apply a specific treatment is clearly beneficial. Conversely, if such a special treatment can be determined, then it opens the door to probing LLP models with \textsc{Contur}.

We propose the following approach to treat LLPs in the \textsc{Rivet} and \textsc{Contur} machinery. The basic idea is to evaluate, for a given set of model parameters, the lab-frame decay lengths for all particles with small widths (and therefore long lifetimes) as determined by the event generator. We can then sort the particles into four categories:
\begin{itemize}
    \item a) those which decayed early enough to be treated as prompt; 
    \item b) neutral or charged particles  which decayed within the detector volume and hence would be unlikely to be picked up by the standard reconstruction algorithms used for measurements if not actively discarded;
    \item c) (heavy) charged particles which decayed outside the detector, but whose interactions in the detector would lead to problematic reconstruction (such as anomalous ionization) which would potentially be vetoed in most measurements;
    
     \item d) neutral particles  which decayed outside the bounds of the detector and which would be reconstructed as missing transverse energy.
\end{itemize}
The individual particles which were not reconstructible can be manually removed from the simulated events (for instance with \textsc{PyHEPMC}~\cite{BUCKLEY2021107310}). Events which contained decays within the detector volume or which would have anomalous interactions with the detector, causing signatures similar to instrumental noise, can be entirely removed.  Once the relevant particles and/or events have been removed from the simulations, the rest of the \textsc{Contur} workflow can then run normally. We have implemented this workflow, and extracted new constraints on LLP models as discussed in Sections~\ref{sec:fimps},~\ref{sec:HS},~\ref{sec:HZZd} and~\ref{sec:ALP}.  This  approach is conceptually related to reweighting-based approaches previously described in \textsc{SModelS}~\cite{Heisig:2018kfq} and in a standalone study of dark radiation using LLP searches~\cite{Bernreuther:2022bdw}, however, it goes further by making event-by-event decisions on which particles to reconstruct, and it is the first time that any such  approach has been used with detector-corrected LHC measurements.
We will now review the step-by-step implementation of our  method, including technical details.

 \paragraph{Step 1} A regular \textsc{Contur} scan on a set of parameters of a model is performed, ensuring that the events (in \textsc{HEPMCs} format) are kept rather than deleted, as they normally would be to save disk space.

 \paragraph{Step 2} At each point in the grid, the generator log files are parsed to determine the list of particle species, in terms of their unique particle data group identifier (\textsc{PDGid}), which have small widths and therefore macroscopic decay lengths. If one already knows which particles are LLPs in the model, one need only save the width of those particles at this stage. But alternatively, one can select all particles with a width less than $\sim 10^{-6} $ eV (for displacements above the millimeter scale).

 \paragraph{Step 3} For the list of LLP \textsc{PDGid}s at each point, we parse the \textsc{HEPMC} files to extract the LLP kinematics for each event, such that time dilation factors can be calculated. Multiple particle species can be long-lived at a given point, and there can be multiple instances of each species in a given event. For each LLP, a random decay length is assigned, sampling from a falling exponential distribution whose mean is the inverse of that particle species' width (as saved in Step 2). The time dilation factor for that particular particle is applied to obtain a lab-frame decay position for each LLP, separately for each event. In principle, this step can be performed by the MCEG programme, but we choose to do it by hand such that we can treat the lifetime as an independent parameter.

 \paragraph{Step 4} We determine a  transverse ``prompt threshold'' $p_{th}$ before which particles can be considered indistinguishable from prompt. Indeed, if the decay of the LLP takes place before $p_{th}$, then tracks and deposits from the decay products should be reconstructible using the standard ATLAS and CMS algorithms. To choose the value of the threshold, we are guided by Refs.~\cite{ATLAS:2023nze} and~\cite{CMS:2014pgm}. The ultimate limiting factor in track or vertex reconstruction is the transverse impact parameter requirement, which is of the order of 5~mm for both experiments. Even in the most extreme cases, where the daughter particles' tracks have a large opening angle, decays before 5~mm would yield reconstructed tracks. We therefore choose $p_{th}$ to be 5~mm. We also determine if particles decay outside of the detector, conservatively taking the dimensions of the ATLAS detector (13~m in radius, and 22~m in length on each side).

 \paragraph{Step 5} For each point of the grid, we parse the \textsc{HEPMC} file using  \textsc{PyHEPMC}. For each event, based on the lifetime and Lorentz boost obtained in previous steps, we determine if each particle decayed a) promptly (with transverse decay position below $p_{th}$), b) in the detector , c) outside of the detector and charged or d) outside the detector and neutral. For particles in case a) nothing happens: the particles are hence treated as if they were prompt. If any LLP (charged or neutral)  decays after the prompt threshold but within the detector, or if a charged particle decays outside the detector (cases b) and c)), the whole event is removed. Finally, in case d), the particle and all its daughters in the decay history in the \textsc{HEPMC} event are removed using the \textsc{PyHEPMC} functionality, with the rest of the event left intact. Events which are removed are still considered when calculating the total number of generated events for normalisation purposes. A new filtered \textsc{HEPMC} is created in this way, for each point in the parameter grid.

 \paragraph{Step 6} We then pass the filtered \textsc{HEPMC} file through \textsc{RIVET} to obtain \textsc{YODA} files in each point of the parameter grid. The rest of the \textsc{Contur} pipeline is executed as normal on the new \textsc{YODA} files.

\vspace{0.5cm}
The effect of this method is to reduce the sensitivity \textsc{Contur} would have reported if no filtering had taken place, since we only ever remove events and/or particles, although we note that one possible exception is if the procedure introduces excessive missing transverse energy, while little would be expected from the theory. We can in principle identify three regimes: first, when LLPs have low lifetimes, the method is effectively extrapolating the prompt sensitivity according to the proportion of reconstructible particles. On the other hand, in the long lifetime regime, neutral LLPs are treated as missing energy, but there can be residual sensitivity to their existence, for example by observing the effect of their recoil on the transverse momentum of any prompt particles which they are produced with. In the intermediate regime, where LLPs decay within the detector, we remove events entirely while conservatively assuming that such events in data could be discarded and hence not be accounted for in the existing measurements. This implies a ``dip" in sensitivity from this method in the intermediate regime for neutral LLPs. Thankfully, this is exactly the regime which direct LLP searches are designed to probe: hence illustrating the nice complementarity of the modified \textsc{Contur} method we propose with the direct search programme.

Since this method takes the same final states that would be predicted by a model, and down-weights the predicted yields, it does not incur any additional model dependence.  Further constraints on LLP models could likely be obtained from the other interactions that are possible in the same model, but do not involve LLPs. A good example is if a pair of LLPs is produced from the decay of a scalar mediator, which can be produced via gluon--gluon fusion, then that same mediator must necessarily be able to decay back into gluons (and with no LLPs involved at all in that reaction). One could then check the standard dijet measurements and searches for the corresponding bump at the scalar mediator mass. This second method, involving only prompt particles, could then use the standard \textsc{Contur} workflow without any special treatment, but it would involve additional assumptions on what is or is not allowed by a particular model. We do not explore this avenue further, and reserve it for future work.

The method outlined in this section has a potential downside. Indeed, in regimes where many events are removed, the results can be subject to large statistical fluctuations. This effect is mitigated in the results which follow by the fact that the statistical uncertainties of the signal samples are accounted for in the \textsc{Contur} machinery when calculating limits. Further, the most stringent (and relevant) limits are obtained in regions where the fraction of kept events is high. However, we note that it is in principle possible to calculate the probability of an event being kept analytically rather than relying on sampling an exponential. We leave this development for future work.

\section{Constraints on feebly interacting DM}\label{sec:fimps}

\begin{figure}[t!]
    \centering
    \includegraphics[width=0.35\linewidth]{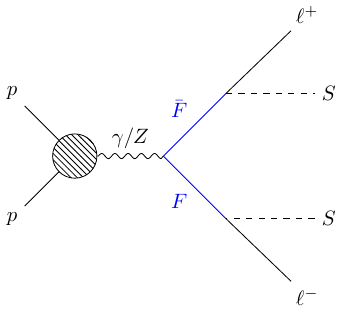}
    \caption{Diagram for the main production process for FIMPs at the LHC, assuming the leptonic version of the model. The blue lines and text indicate the long-lived particle(s).}
    
    \label{fig:fimp-diagram}
\end{figure}
To illustrate how the method described in Section~\ref{sec:contur_for_llps} can be used to obtain new constraints on LLP models, we first consider a simple freeze-in model explaining the observed DM abundance through the decay of an electrically- and/or colour-charged long-lived parent particle into a Feebly Interacting Massive Particle (FIMP), as proposed in Contribution 6 of Ref.~\cite{Brooijmans:2020yij}, described in more detail in Ref.~\cite{Belanger:2018sti}, and also studied in~\cite{Bernreuther:2022bdw}. Freeze-in is an alternative to the frozen-out Weakly Interacting Massive Particle (WIMP) picture which, despite its attractive features, is under some tension with null experimental evidence. This FIMP model introduces a real scalar DM candidate, denoted $S$, which is neutral under the SM gauge group. Additionally, a vector-like fermion $F$ plays the role of mediator. In this model, $F$ is electrically charged and can acquire a long lifetime if the coupling governing its decay is small. We consider here the leptonic version of this minimal freeze-in model, where the DM candidate is coupled to the SM through Yukawa-type terms involving the vector-like fermion and the SM charged leptons. The LHC signature of the leptonic model is then the Drell--Yan pair-production of the vector-like fermion (with electric charge $\pm 1$), followed by its decay into the scalar DM candidate along with a charged lepton, as shown in Figure~\ref{fig:fimp-diagram}. This last decay can be displaced or even take place outside the detector. Another version of the model, where $F$ decays hadronically, exists but is not studied further here. 

This FIMP model was confronted with LHC search results in Ref.~\cite{Brooijmans:2020yij,Belanger:2018sti} for different assumptions of LLP lifetime: an ATLAS search for prompt leptons and missing transverse momentum~\cite{ATLAS:2019lff} was used for short lifetimes, CMS searches for displaced leptons~\cite{CMS:2016isf} or displaced tracks~\cite{CMS:2018rea} were used for intermediate lifetimes, and an ATLAS search for heavy stable charged particles~\cite{ATLAS:2019gqq} was used for detector-stable LLPs. Together, these searches provided coverage for $c\tau$ values ranging from tens of micrometers to kilometers, excluding LLP masses between about 150 GeV and 350-700 GeV depending on the lifetime. A different set of re-interpretations for the same model are also available in Ref.~\cite{Bernreuther:2022bdw}.

To determine any additional constraints that could be extracted from existing measurements using \textsc{Contur}, we simulated the process encoded in the publicly available UFO file for the model~\cite{FIMP_UFO} using  \textsc{MadGraph5\_aMC@NLO} v3.4.2~\cite{Alwall:2014hca} and \textsc{Pythia} v8~\cite{Sjostrand:2007gs} for hadronisation. For each point, 10~000 events were generated. The grid was composed of LLP masses in the range of 100 to 1000 GeV in steps of 45 GeV. The lifetime of the $F$ was assumed to be independent of its mass, and 20 values of $c\tau$ were scanned on a logarithmic scale between 0.01~mm and 10~m. The mass of the DM candidate was set to 12 keV, although the analysis presented here is not strongly sensitive to this choice. Since the LLP $F$ is electrically charged, it would in principle form a reconstructible track before its decay, although this could have anomalous ionization and hence might not be reconstructed properly. After the decay, a SM charged lepton may continue the track, but since the unseen $s$ is emitted, one would expect the track from the $F$/$\ell$ to be ``kinked'': in other words for the hits not to be aligned in the expected pattern for a charged particle trajectory, but instead be the discontinuous union of two curved tracks. Such signatures would be missed by ATLAS and CMS, as standard tracking algorithms look for a single particle trajectory. Both these consideration motivate the need to veto events where the decay takes place after the prompt threshold. If the decay takes place before the prompt threshold discussed in Section~\ref{sec:contur_for_llps}, the lepton track would still be reconstructed as normal. 
%For both ATLAS and CMS, the inner tracking system's penultimate layer is at around 10~cm in the barrel and around 50~cm in the endcaps, so our default e LLP un-reconstructable.
We then follow the procedure set out previously to determine the exclusions from \textsc{Contur}. 
\begin{figure}[t!]
    \centering
%\subfloat[   
    \includegraphics[width=0.5\linewidth]{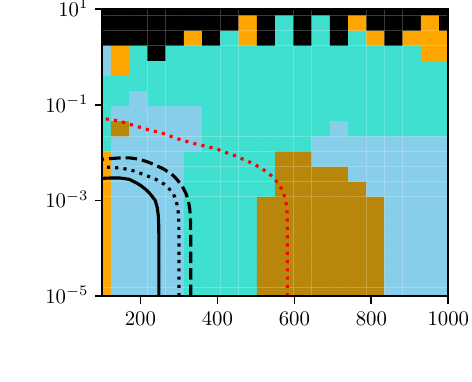}
    \begin{minipage}{0.45\textwidth}
    \vspace{-6.5cm}
\includegraphics[width=0.98\linewidth]{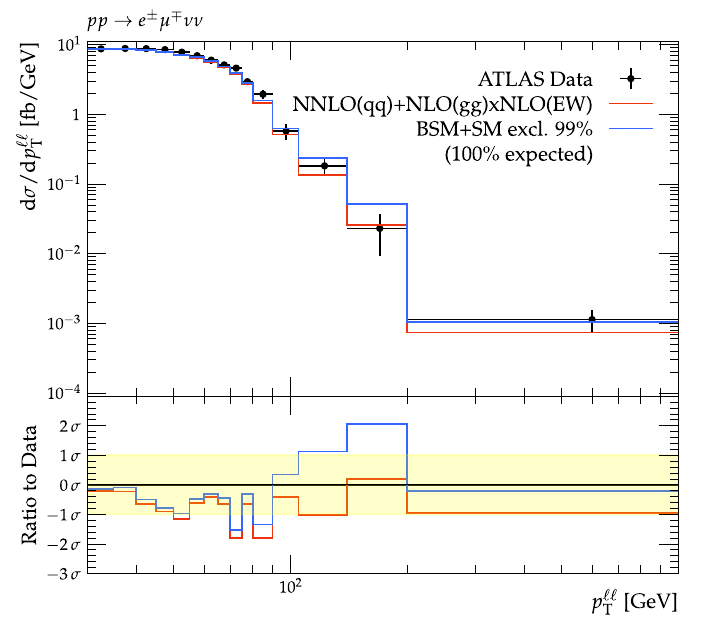}   
 \end{minipage} 
 %  \begin{flushleft}
 %   \begin{minipage}{.45\linewidth}  
 %  \begin{tabular}{llll}
 %  \quad    & \swatch{darkgoldenrod}~$\ell^+\ell^-\gamma$  & 
 %       \swatch{skyblue}~$\ell_1\ell_2$+\MET{} \cite{ATLAS:2019rob} & %\\ 
 %   \quad  &  \swatch{turquoise}~$\ell_1\ell_2$+\MET{}+jet %\cite{ATLAS:2021jgw} &  \end{tabular}    
 % \end{minipage} 
 %   \end{flushleft} 

    \begin{minipage}{.9\linewidth}  
  
   \begin{tabular}{llll}
        \swatch{black}~ No data  &
        \swatch{turquoise}~$\ell_1\ell_2$+\MET{}+jet \cite{ATLAS:2021jgw} &
        \swatch{skyblue}~$\ell_1\ell_2$+\MET{} \cite{ATLAS:2019rob} &
        \swatch{blue}~$\ell$+\MET{}+jet  \\
        \swatch{orange}~$\ell^+\ell^-$+jet \cite{ATLAS:2017nei} &
        \swatch{darkgoldenrod}~$\ell^+\ell^-\gamma$  &
   \end{tabular}
  \end{minipage} 
    \caption{(Left) Exclusion contours for the FIMP model as a function of the $F$ mass and $c\tau$, assuming $m_S=12$ keV. The solid (dotted) lines represent the observed (expected) exclusions at 95\% CL. The dashed line represents the observed 68\% CL exclusion. The red dotted line represents the 95\% CL expected exclusion extrapolated to the HL-LHC. The area below the curves is excluded. The colours of the cells represent the dominant analysis pool for each point in the grid. (Right) The dominant exclusion histogram for a particular excluded point, where $m_F=147$ GeV and $c\tau=0.0008$~m. The exclusion comes from the dilepton invariant mass spectrum measured in the $e\mu\nu\nu$ channel of an ATLAS $W^+W^-$ measurement~\cite{ATLAS:2019rob}, specifically, Figure 7c of that paper. The black markers represent the observed data, and the red line shows the SM prediction. The blue line shows the SM prediction summed with the new physics prediction, which in this case results in a significant disagreement with the observation.}
    \label{fig:FIMP-results}
\end{figure}
\begin{figure}[h!]
    \centering
   \hspace{2cm} \includegraphics[width=0.6\linewidth]{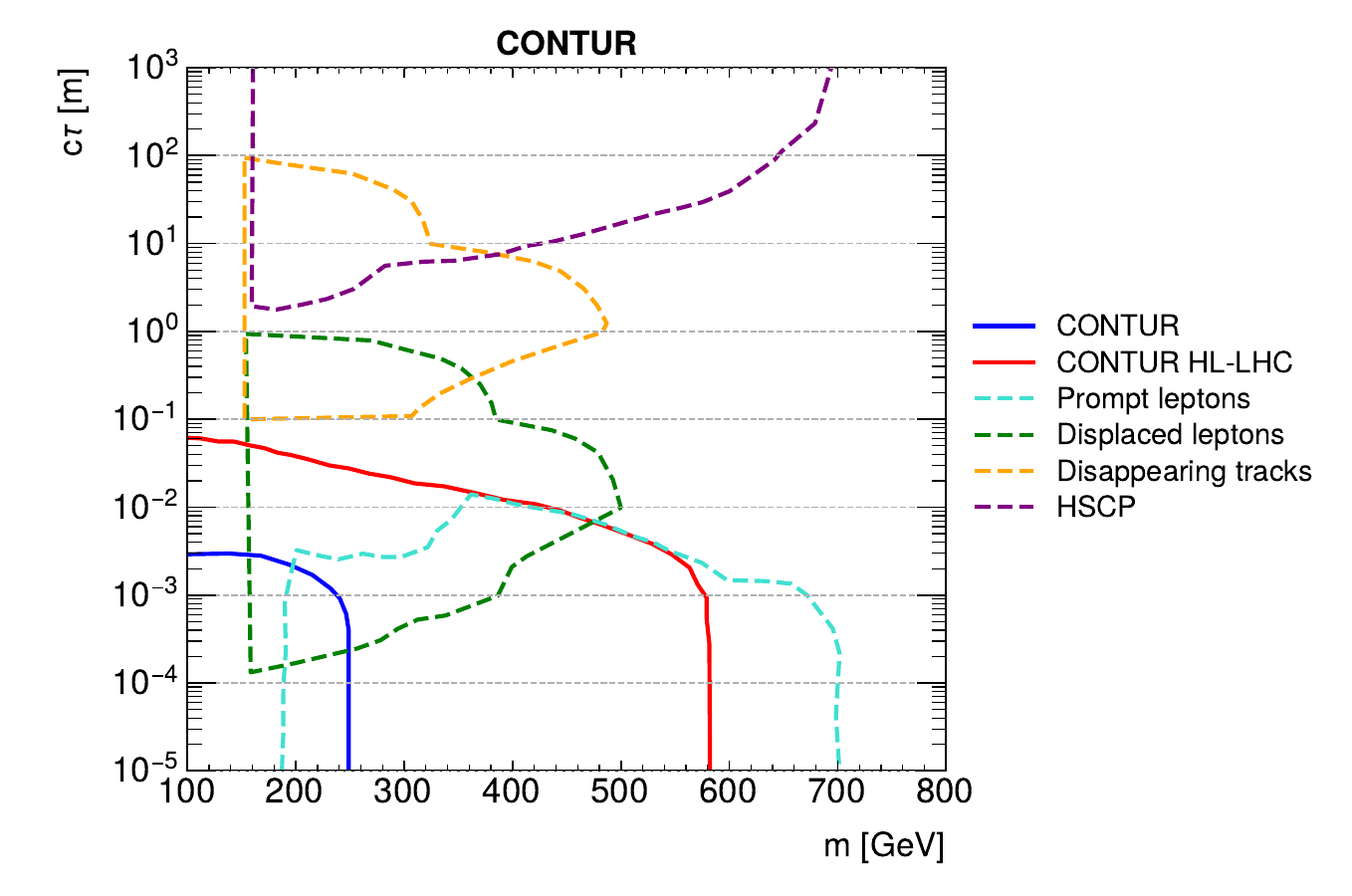}\\

    \caption{ Comparison of \textsc{Contur} results with current measurements, extrapolated \textsc{Contur} results for HL-LHC, and recasted search results for the FIMP model from Ref.~\cite{Brooijmans:2020yij}.}
    \label{fig:FIMP-results-summary}
\end{figure}

The results are shown in Figure~\ref{fig:FIMP-results}, indicating that $F$ masses from 100 GeV up to $\sim$250 GeV can be excluded with 95\% confidence level (CL) so long as the mean proper lifetime times the speed of light is below $\sim$2~mm. The colours in the figure represent the type of measurement that brings the exclusion power: for masses below about 300~GeV, the exclusion is dominated by analyses that produce two (different-flavour) leptons and missing transverse momentum. For higher masses, other final states dominate the exclusion, notably opposite-flavour leptons with missing transverse energy and the presence of one of more jets, or alternatively measurements of dilepton production with an additional photon. Focusing on the excluded region, taking a point where the mass of $F$ is approximately 150 GeV, Figure~\ref{fig:FIMP-results} also shows the exclusion histogram from the ATLAS measurement of $W^+W^-$ production~\cite{ATLAS:2019rob} that generated the highest exclusion. Indeed, the presence of such a signal would have significantly altered the predicted the shape of the spectrum.

In Figure~\ref{fig:FIMP-results-summary}, we compare the results obtained from \textsc{Contur} to limits derived from direct searches in Ref.~\cite{Brooijmans:2020yij}. In this case, the direct searches provide more stringent exclusions at high LLP masses, which is perhaps not surprising given that they were originally searches for high-mass supersymmetric particles, and would be more optimal in such a regime. However, these searches have a cut-off around 150 GeV below which they do not set exclusions, although we note that this region is filled in by 8 TeV direct searches for prompt particles~\cite{ATLAS:2018ojr,ATLAS:2014zve}. That region is also covered by the \textsc{Contur} exclusion lines. 
%Further, the projection of these exclusions to the HL-LHC indicate that even without new searches, the limits on the FIMP model could be extended up to around 550 GeV and to $c\tau$ up to around 7~cm, if the same measurements are repeated during the HL-LHC era. 

\section{Constraints on hidden sector models with heavy mediators decaying into long-lived scalars}\label{sec:HS}

\begin{figure}[t!]
    \centering
    \includegraphics[width=0.35\linewidth]{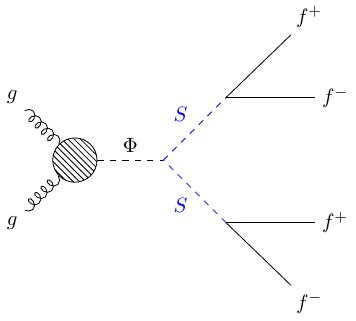}
    %\hspace{2cm}
  %  \includegraphics[width=0.35\linewidth]{figures/FeynmanDiagrams/HAHMZdiagram.pdf}
    \caption{Diagram for  gluon--gluon fusion
    %nd (right) associated $Z$ boson p
    production of a HS mediator $\Phi$ at the LHC, assuming decays into neutral scalars $S$, which later decay into fermions $f$ via a Yukawa-type interaction. In this paper, the $S$ decays chiefly into $b$-quark or $t$-quark pairs if kinematically allowed.}
    \label{fig:hiddensector-diagram}
\end{figure}

Hidden sector models~\cite{Strassler:2006im,Strassler:2006ri,Chan:2011aa,Curtin:2013fra} postulate that the SM is connected to a dark sector of new particles via some scalar mediator $\Phi$, which can be the Higgs boson or some new boson. If the mediator is a scalar, it can be produced (amongst others) via gluon--gluon fusion (gg$\Phi$) just like the Higgs boson. It can decay into a pair of neutral scalars from the dark sector, denoted $S$, which would eventually decay back into the SM via a Yukawa interaction, as shown in Figure~\ref{fig:hiddensector-diagram}. The $S$ can become long-lived because its decay back into the SM occurs through small couplings. The Hidden Abelian Higgs Model (HAHM)~\cite{Wells:2008xg,Curtin:2013fra} can predict such a phenomenology. The HAHM introduces two new scalars $\Phi$ and $S$, and a dark photon $Z_d$. In this section, we focus on the decays of the $\Phi$ to $S$ only (as done by ATLAS).

Generic HS models have been studied by both ATLAS and CMS, for example in recent searches for displaced activity in the tracking system~\cite{ATLAS:2024qoo}, calorimeters~\cite{ATLAS:2024ocv} or muon system~\cite{ATLAS:2022gbw}, ensuring a coverage from  millimeters to tens of meters in c$\tau$ of the LLP $S$, and covering many scenarios for the masses of the particles. Phenomenological re-interpretations of other searches into this model have also been performed, for example in Ref.~\cite{Wang:2024ieo}. Summaries of experimental constraints are available for ATLAS~\cite{ATL-PHYS-PUB-2022-007} and CMS~\cite{CMSsummar}. CMS has mostly focused on the case where the mediator is the Higgs boson, while ATLAS has also considered heavier mediators, up to 1 TeV in mass. 
%For both experiments, the searches at low lifetimes have studied the $V\Phi$ production mode, exploiting the vector boson for triggering and background reduction. 
%Those searches focused on the Higgs-boson-mediated case. 
For those higher mediator masses, no constraints are reported for $c\tau$ below the centimeter range. We attempt to address this lack of coverage using the modified \textsc{Contur} workflow.
\begin{figure}[t!]
    \centering
    \includegraphics[width=0.5\linewidth]{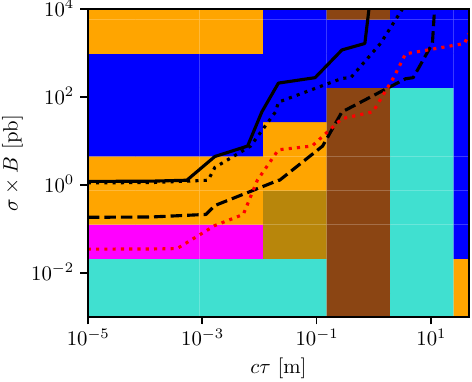}
        \begin{minipage}{0.45\textwidth}
    \vspace{-6.5cm}
\includegraphics[width=0.98\linewidth]{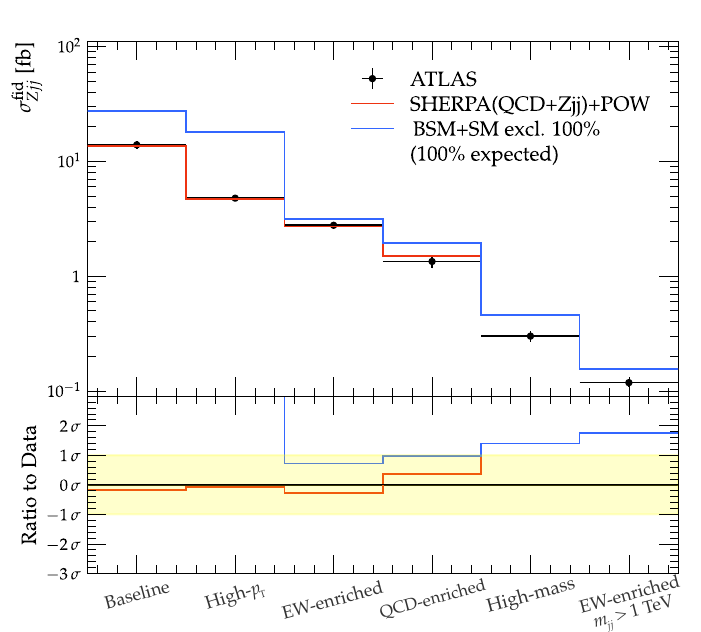}   
\end{minipage}
    \begin{minipage}{.9\linewidth}  
    \hspace{0cm}
    \begin{tabular}{llll}
        \swatch{turquoise}~$\ell_1\ell_2$+\MET{}+jet \cite{ATLAS:2019hau,ATLAS:2019ebv} &
        \swatch{magenta}~4$\ell$ \cite{ATLAS:2021kog} & 
        \swatch{orange}~$\ell^+\ell^-$+jet  \cite{ATLAS:2017nei,ATLAS:2020juj}  \\
        \swatch{blue}~$\ell$+\MET{}+jet \cite{CMS:2018htd,ATLAS:2017cez,CMS:2016oae,ATLAS:2019hxz} &
        \swatch{saddlebrown}~hadronic $t\bar{t}$ \cite{ATLAS:2020ccu,CMS:2019eih} &
        \swatch{darkgoldenrod}~$\ell^+\ell^-\gamma$ \cite{ATLAS:2019gey} &
    \end{tabular}  
  \end{minipage} 
   \caption{(Left) Exclusion contours for the HS model as a function of cross-section times branching fraction of $gg\Phi(SS)$ and $c\tau$, assuming $m_S=475$ GeV. The solid (dotted) lines represent the observed (expected) exclusions at 95\% CL. The dashed line represents the observed 68\% CL exclusion. The red dotted line represents the 95\% CL expected exclusion extrapolated to the HL-LHC. The area above the curves is excluded. The colours of the cells represent the dominant analysis pool for each point in the grid. (Right) The dominant exclusion histogram for a particular excluded point, where the cross-section is around 2.15~pb and $c\tau=0.00031$~m. The exclusion comes from an ATLAS measurement of the cross-section for electroweak production of dijets in association with a $Z$ boson~\cite{ATLAS:2017nei}, specifically, Table 3 of that paper. The black markers represent the observed data, and the red line shows the SM predictions. The blue line shows the SM prediction summed with the new physics prediction, which in this case results in a significant disagreement with the observation.}
    \label{fig:HAHM_475-results}
\end{figure}
We focus on the case of a heavy mediator of mass $1000$ GeV produced via gg$\Phi$ production only.
The signature consists of up to four $b$-jets or products of top quark decays if the LLP mass is above the relevant kinematic threshold. The events are generated with \textsc{MadGraph5\_aMC@NLO} v3.4.2 and \textsc{Pythia} v8 for hadronisation, using the existing UFO implementation of the HAHM~\cite{HAHM_UFO} to establish direct comparisons to direct searches. The coupling of the dark photon to the SM and mediator is set to zero. 
Several values of the neutral scalar $S$ mass are considered, between 150 and 475 GeV, to match the parameter values chosen by ATLAS. 
%The $Z$ boson is allowed to decay into electrons, muons or taus.
For each assumption, the grid scan consisted of 10 values of $c\tau$ on a logarithmic scale between 0.01~mm and 100~m, and 10 values of cross-section times branching fraction ($\sigma \times B$) on a logarithmic scale between 0.001~pb and 1000~pb, with 10000 events generated at each point. We note here the high end of the considered cross-sections could be in a non-perturbative regime, but we ignore this effect in this study. The kinematics of the events are assumed to be uncorrelated to the LLP mean proper lifetime, as was done for the ATLAS searches.

\begin{figure}[t!]
    \centering
    \includegraphics[width=0.5\linewidth]{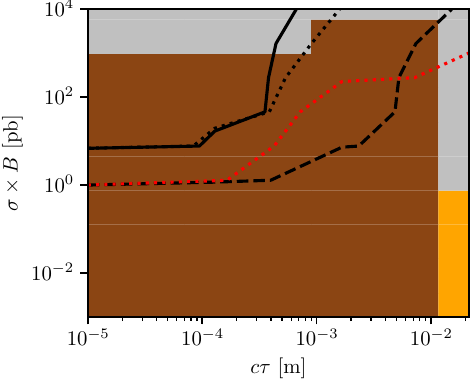}
        \begin{minipage}{0.45\textwidth}
    \vspace{-6.5cm}
\includegraphics[width=0.98\linewidth]{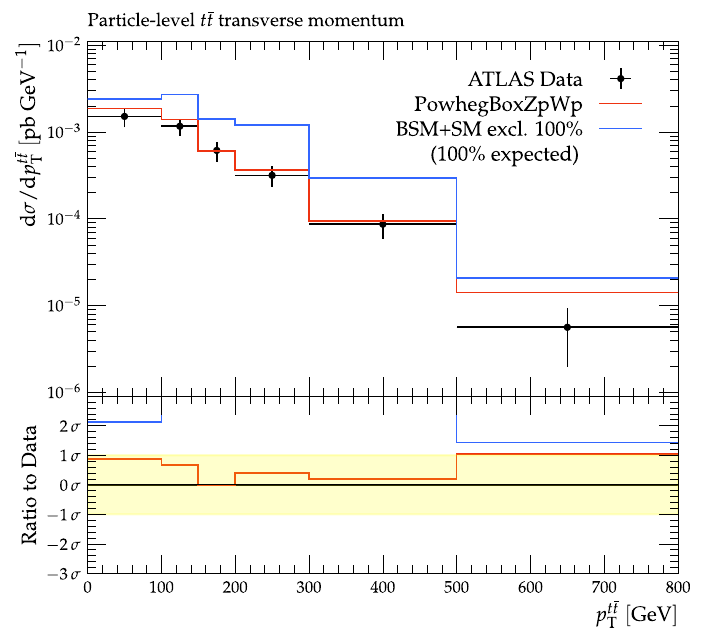}   
\end{minipage}
    \begin{minipage}{.9\linewidth}  
    \hspace{3cm}
    \begin{tabular}{llll}
    \swatch{silver}~jets \cite{ATLAS:2017ble} &
        \swatch{saddlebrown}~hadronic $t\bar{t}$ \cite{ATLAS:2018orx} & 
        \swatch{orange}~$\ell^+\ell^-$+jet  & 
    \end{tabular}
  \end{minipage} 
   \caption{(Left) Exclusion contours for the HS model as a function of cross-section times branching fraction of $gg\Phi(SS)$ and $c\tau$, assuming $m_S=150$ GeV. The solid (dotted) lines represent the observed (expected) exclusions at 95\% CL. The dashed line represents the observed 68\% CL exclusion. The red dotted line represents the 95\% CL expected exclusion extrapolated to the HL-LHC. The area above the curves is excluded. The colours of the cells represent the dominant analysis pool for each point in the grid. (Right) The dominant exclusion histogram for a particular excluded point, where the $\sigma\times B$ is around 10~pb and $c\tau=0.0001$~m. The exclusion comes from an ATLAS measurement of boosted $t\bar{t}$ production~\cite{ATLAS:2017nei}, specifically, Figure 4a of the auxiliary material of Ref.~\cite{ATLASttbarAuxMat}. The black markers represent the observed data, and the red line shows the SM predictions. The blue line shows the SM prediction summed with the new physics prediction, resulting in a significant disagreement with the data.   }
    \label{fig:HAHM_150-results}
\end{figure}

The workflow described in Section~\ref{sec:contur_for_llps} is exploited to determine what constraints can be extracted from the bank of LHC measurements for HS models with heavy mediators. Since the displaced hadronic activity from LLPs would lead to anomalous jets, which would be trackless and therefore vetoed by the standard experimental reconstruction algorithms, we maintain the prompt threshold values described previously. An example of the results for the case where the $S$ mass is set to 475 GeV is shown in Figure~\ref{fig:HAHM_475-results}. In this case, one expects the LLP to decay mostly into top quarks.  Cross-sections below around 1~pb are excluded up to 1~mm in $c\tau$. 
The exclusion is mostly driven by 13 TeV ATLAS measurements involving one lepton, missing transverse momentum and jets or measurements of pairs of leptons. Figure~\ref{fig:HAHM_475-results} focuses on one of the excluded points: the existence of this signal would have caused divergences in the Baseline and High-$p_\textrm{T}$ bins of the measurement presented in Ref~\cite{ATLAS:2017nei}. 
Figure~\ref{fig:HAHM_150-results} shows equivalent information for a case where the LLP mass is below the top-quark threshold, such that most of the decays are into $b$-quarks. In this case, the excluded cross-sections peak at around 10~pb for nearly-prompt particles before dropping off after around 0.2~mm.  The dominant analysis is a measurement of $t\bar{t}$ production~\cite{ATLAS:2018orx}, which would indeed yield the same final state of a pair of leptons accompanied by $b$-jets. In neither example do we perceive any exclusion for the high-lifetime regime, which makes sense because, at leading order the LLPs are produced without anything to recoil against, and hence their presence can only inferred if the LLPs themselves are detectable.
\begin{figure}[t!]
    \centering
\includegraphics[width=0.78\linewidth]{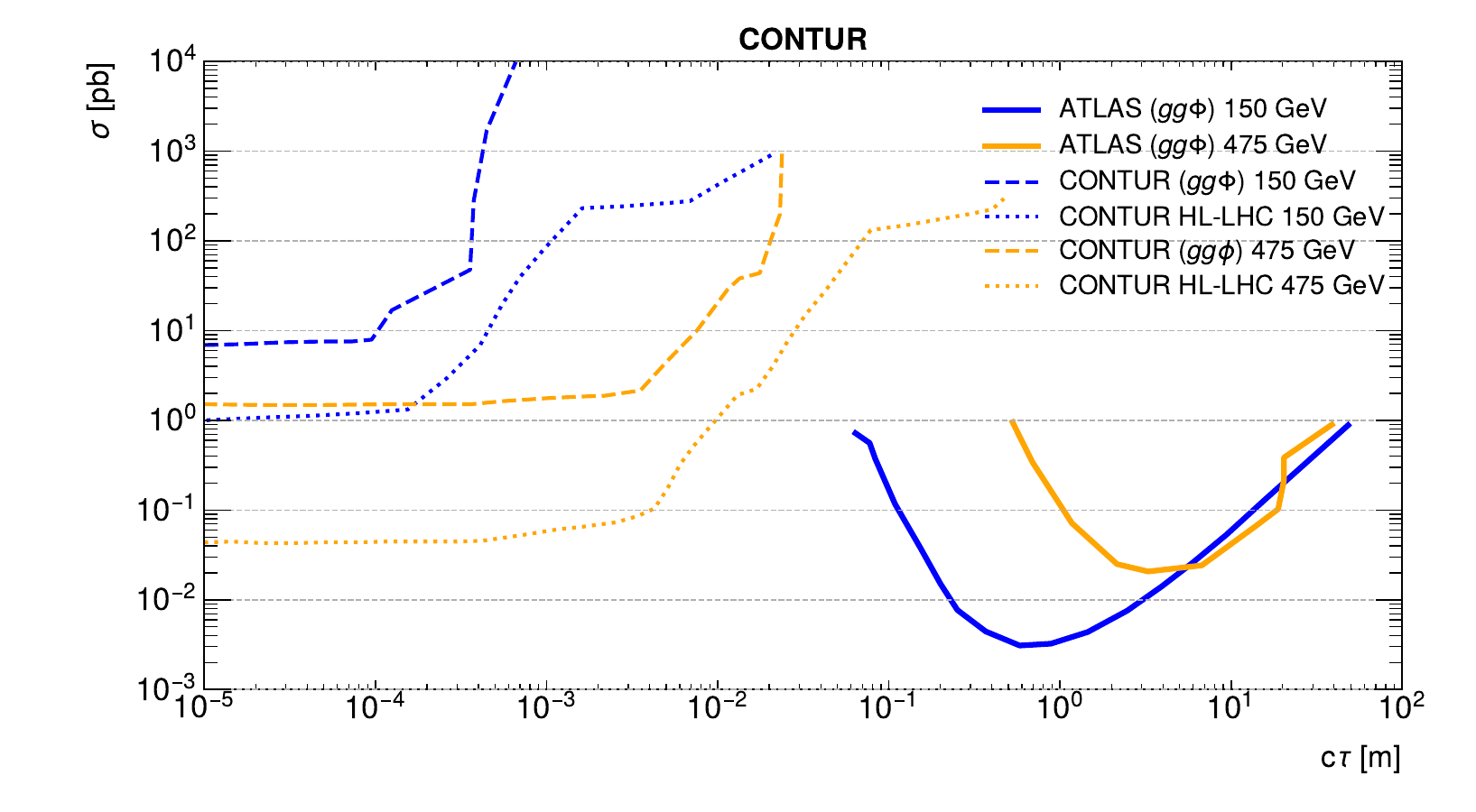}\\
 \includegraphics[width=0.78\linewidth]{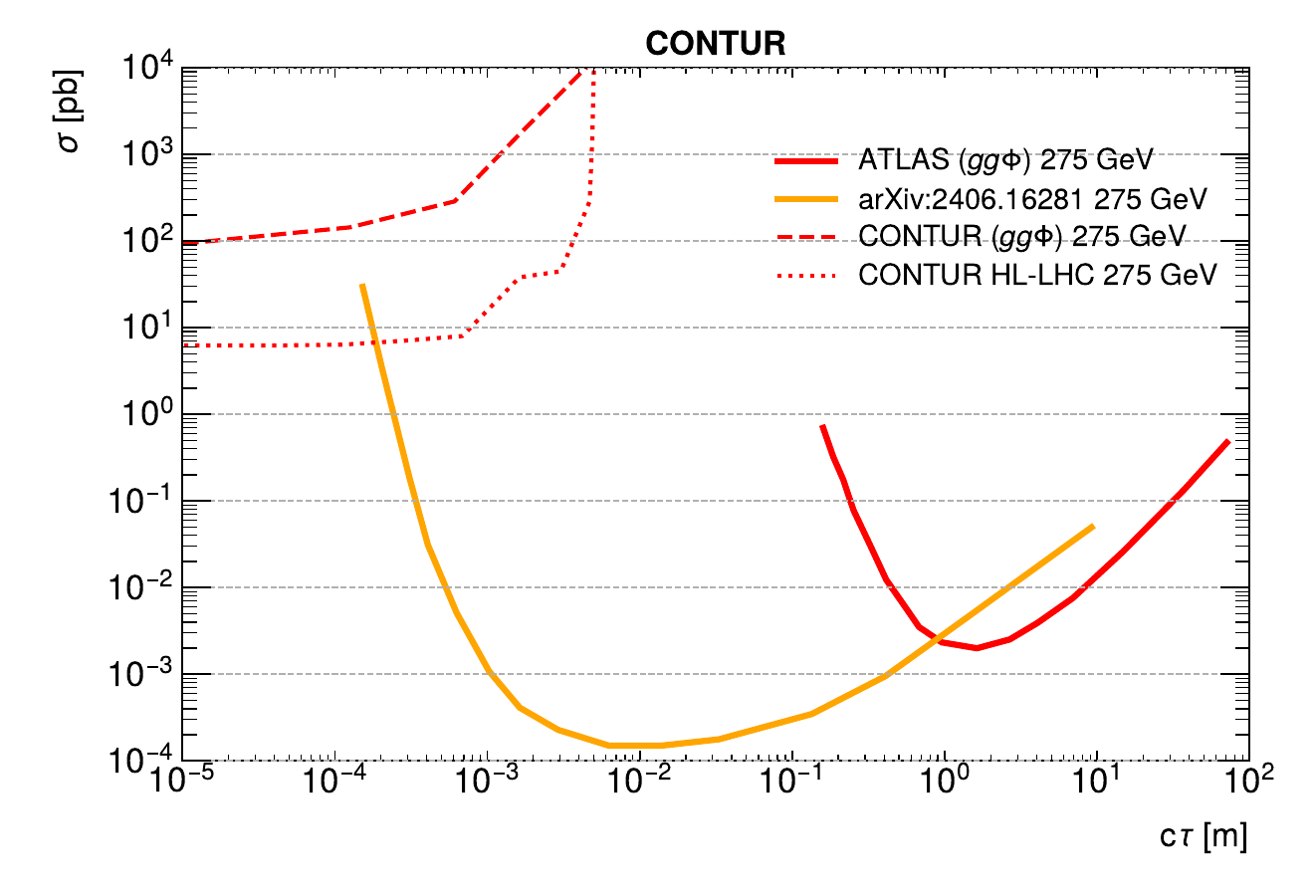}
    \caption{Comparison of \textsc{Contur} results for the HS model for a mediator of mass 1 TeV with current measurements, extrapolated \textsc{Contur} results for HL-LHC, and comparable results from an ATLAS search~\cite{ATLAS:2024ocv} and a phenomenological re-interpretation study~\cite{Wang:2024ieo} for the same model and parameter choices.}
    \label{fig:HAHM-results-summary}
\end{figure}

The comparison of the results using \textsc{Contur} with the best results from ATLAS is shown in Figure~\ref{fig:HAHM-results-summary}. The regions excluded by \textsc{Contur} are in regions where no constraints from direct searches exist so far. Although the \textsc{Contur} results are comparatively weak, they provide complementary constraints in a region which the direct searches have not yet explored.

\section{Constraints on heavy, long-lived dark photon models}\label{sec:HZZd}

\begin{figure}[t!]
    \centering
    \includegraphics[width=0.35\linewidth]{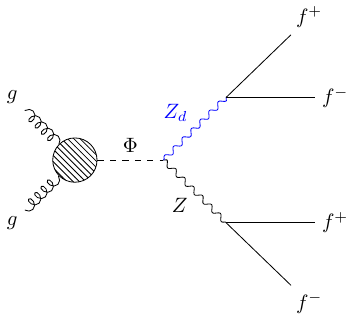}
    \caption{Diagram for the production of a dark photon $Z_d$ via a scalar mediator $\Phi$ at the LHC. The blue lines and text indicate the long-lived particle(s).}
    
    \label{fig:ZD-diagram}
\end{figure}

We now consider a model where a dark photon $Z_d$ is the long-lived particle, as in Refs.~\cite{Curtin:2014cca,Davoudiasl:2012ag}. In this model, the  $Z_d$ is produced in association with a regular $Z$ boson from the decay of the mediator $\Phi$ described in Section~\ref{sec:HS}. It acquires a long lifetime through suppressed couplings to the SM, and decays mostly into quark--antiquark pairs or pairs of leptons. A Feynman diagram showing the processes under study is shown in Figure~\ref{fig:ZD-diagram}.
This model was studied by ATLAS in Refs.~\cite{ATLAS:2018niw,ATLAS:2024ocv}. The most constraining results from those searches restrict cross-sections to be smaller than around 0.1~pb in the 10~cm to 10~m range, but no constraints exist so far for $c\tau$ values below 10~cm from ATLAS or CMS. Further, the limits from the direct searches weaken considerably for proper lifetimes above 100~m or so, as the detectors lose acceptance for LLP decays.

Events were generated with \textsc{MadGraph5\_aMC@NLO} v3.4.2 and \textsc{Pythia} v8 for hadronisation, starting from the HAHM UFO~\cite{HAHM_UFO} file used in Section~\ref{sec:HS} but turning the couplings of the mediator $\Phi$ to $S$ particles to zero and setting the $S$ mass to a very high scale. The grid is composed of 10 points spaced logarithmically in $\sigma \times B$ between 0.01~mm and 10~m and 10 points spaced logarithmically in $c\tau$ between 1~mm and 1~km. The cross-section and kinematics of the model are assumed to be independent of the $c\tau$, with 10~000 events generated for each combination. Several options of the mediator mass between 400 and 600 GeV, and the dark photon mass between 50 and 400 GeV, are considered, to match those probed by the ATLAS experiment. 

\begin{figure}[t!]
    \centering
       \hspace{-0.75cm}
    \includegraphics[width=0.465\linewidth]{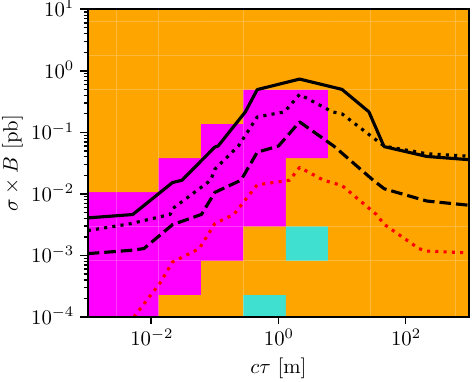}
     \begin{minipage}{0.45\textwidth}
     \vspace{-6cm}
     \hspace{1cm}
    \begin{tabular}{l}
        \swatch{turquoise}~$\ell_1\ell_2$+\MET{}+jet \cite{ATLAS:2019hau} \\
        \swatch{orange}~$\ell^+\ell^-$+jet \cite{ATLAS:2017nei} \\
        \swatch{magenta}~4$\ell$ \cite{ATLAS:2021kog} \\
    \end{tabular}
     \end{minipage}
       \includegraphics[width=0.45\linewidth]{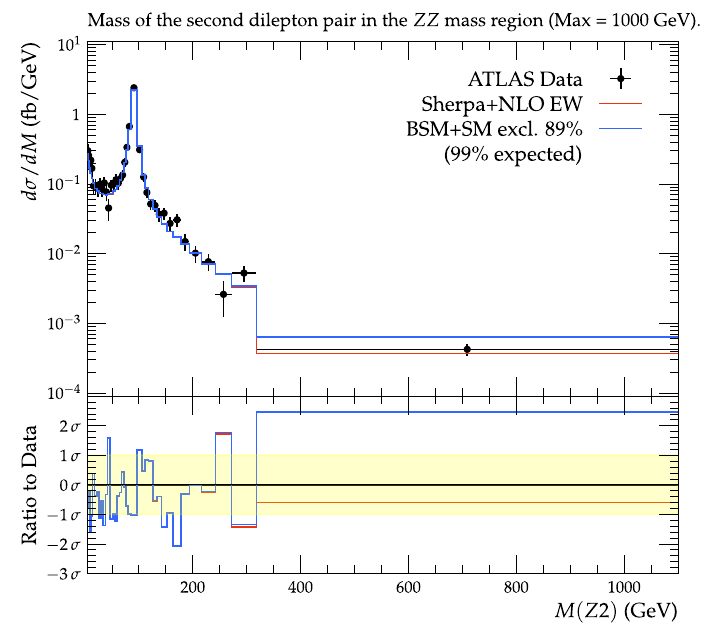}
        \includegraphics[width=0.45\linewidth]{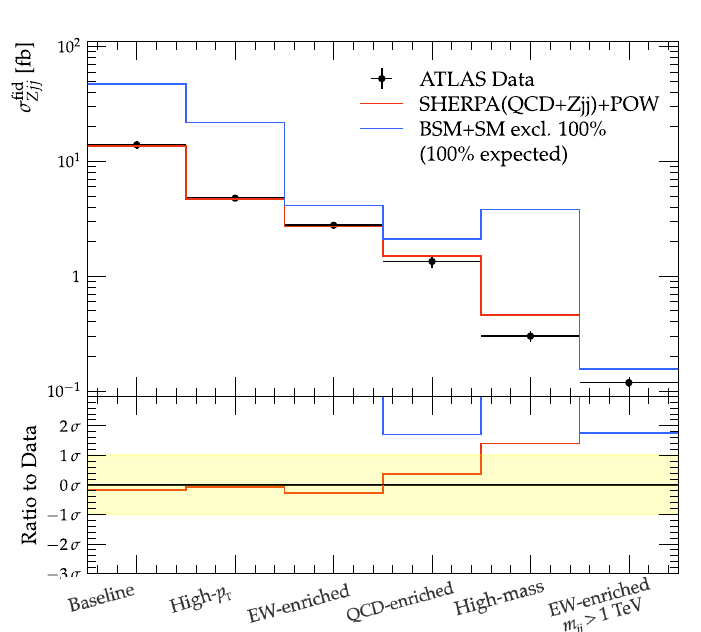}

    \caption{(Top) Exclusion contours for the dark photon model as a function cross-section times branching fraction and $c\tau$, assuming $\Phi$ has a mass of 600 GeV and $Z_d$ has a mass of 400 GeV. The solid (dotted) lines represent the observed (expected) exclusions at 95\% CL. The dashed line represents the observed 68\% CL exclusion. The red dotted line represents the 95\% CL expected exclusion extrapolated to the HL-LHC. The area above the curves is excluded. The colours of the cells represent the dominant analysis pool for each point in the grid. (Bottom left) The dominant exclusion histogram for $\sigma\times B = 0.1$~pb and $c\tau=5$~mm. The exclusion comes from a measurement of the four-lepton invariant mass spectrum~\cite{ATLAS:2021kog}, specifically, from the measurement of the mass of the second lepton pair in Figure 7d of that paper. (Bottom right) The dominant exclusion histogram fore $\sigma\times B = 0.2$~pb and $c\tau=1000$~m. The exclusion comes from an ATLAS measurement of the cross-section for electroweak production of dijets in association with a $Z$ boson~\cite{ATLAS:2017nei}, specifically, Table 3 of that paper.
    The black markers represent the observed data, corrected for detector effects, and the red line shows the SM predictions. The blue line shows the SM prediction summed with the new physics prediction, which in this case results in a significant disagreement with the observed data in the last bin.}
    \label{fig:Zd-results}
\end{figure}

The results of the \textsc{Contur} study are presented in Figure~\ref{fig:Zd-results} as a function of $c\tau$ and cross-section times branching fraction, for the case where the mediator mass is 600 GeV and the dark photon mass is 400 GeV.
In this figure, the three exclusion regimes can be seen clearly. For low lifetimes, the model can be excluded down to the 0.01~pb level, exploiting measurements of the four-lepton invariant mass and dilepton+jets measurements. The sensitivity starts to decrease for $c\tau$ values around 1~cm, in a way that is complementary to the ATLAS search programme. In the middle lifetime regime, the sensitivity degrades since many particles decay within the detector volume and hence the events are ignored. However, at higher lifetimes, once the LLPs start to decay chiefly outside of the detector, sensitivity is regained: indeed, in such cases the presence of the LLP  can be inferred from a recoil-induced bump in the transverse momentum spectrum of the associated $Z$ boson. This long-lifetime sensitivity is independent of $c\tau$ and of the decay mode of the LLP. The red dotted line indicates the sensitivity that could be obtained if the measurements were repeated at the high-luminosity LHC, with constraints down to 0.001~pb possible. Figure~\ref{fig:Zd-results} also shows an example of a histogram from the four-lepton invariant mass measurement~\cite{ATLAS:2021kog}, where the invariant mass of the second lepton pair further away from the $Z$ boson mass shows a large excess compared with the measured data. So, this model can be excluded by the fact that the dark photon can delay leptonically sufficiently often that it would lead to differences in this spectrum beyond what was measured by ATLAS.

\begin{figure}[t!]
    \centering
      \hspace{2cm}  \includegraphics[width=0.75\linewidth]{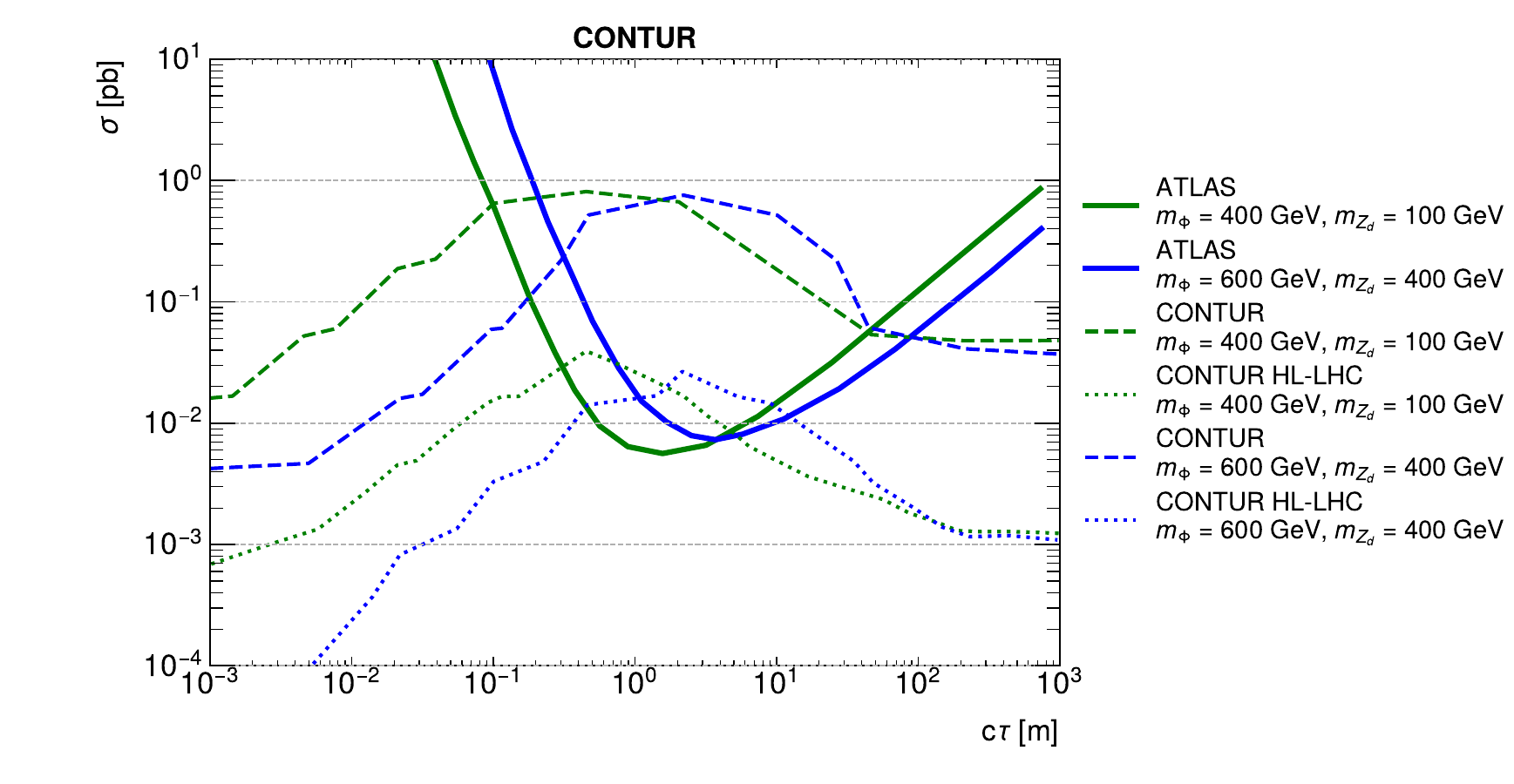}\\
    \caption{ Comparison of \textsc{Contur} results for the $Z_d$ model, extrapolated \textsc{Contur} results for HL-LHC, and the results of a direct search from ATLAS~\cite{ATLAS:2024ocv},}
    \label{fig:Zd-results-summary}
\end{figure}

A comparison of the \textsc{Contur} results for various mass hypotheses against the results obtained by ATLAS in direct searches can be seen in Figure~\ref{fig:Zd-results-summary}. In this model, the \textsc{Contur} results are complementary to the direct searches, effectively ruling out a region not yet covered by the search programme at a comparable cross-section. Further, measurements at the HL-LHC could even eventually exclude the region covered by the direct searches. Hence, it could be concluded that no direct searches are needed for this model in the very low and very high lifetime regimes.

\section{Constraints on long-lived photophobic axion-like particles}\label{sec:ALP}

\begin{figure}[t!]
    \centering
    \includegraphics[width=0.35\linewidth]{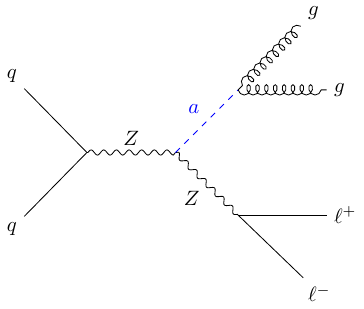}
    \caption{Diagram for the production of an axion-like particle $a$ in association with a $Z$ boson. We consider the case where $a$ decays exclusively into gluons. The blue lines and text indicate the long-lived particle(s).}
    \label{fig:ALP-diagram}
\end{figure}

The final model we consider is a ``photo-phobic'' axion-like particle (ALP) model where the ALPs couple exclusively to gluons (and so decay into jets), as proposed for example in Ref.~\cite{Brivio:2017ije}. In some parts of the parameter space, the ALP can become long-lived due to small couplings mediating the decay. We consider a case where the ALP is produced via associated $Z$ boson production, as shown in Figure~\ref{fig:ALP-diagram}. This model was studied by ATLAS in Refs.~\cite{ATLAS:2024qoo,ATLAS:2024ocv} exploiting decays of the ALP in the tracker and calorimeters respectively. The tracker-based search considered ALPs with masses of 40 GeV and 55 GeV, and set constraints at the level of 0.1~pb in the millimeter to 10~cm range. The calorimeter-based search considered ALPs in masses ranging from 100 MeV to 40 GeV, giving constraints for the 40~GeV mass point in the 10~cm to 10~m range. Lighter ALP mass choices give similar constraints but shifted towards lower $c\tau$ values due to increased time dilation factors. In both cases, the sub-millimeter range however is not well covered.

\begin{figure}[t!]
    \centering
    \hspace{-0.65cm}
    \includegraphics[width=0.465\linewidth]{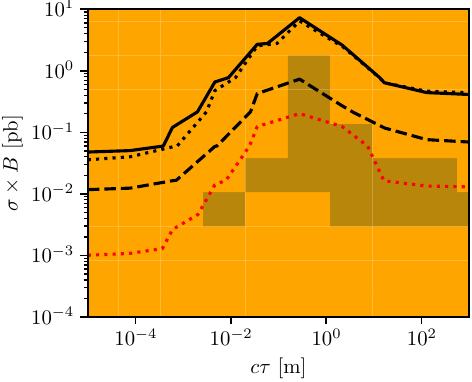}
     \begin{minipage}{0.45\textwidth}
     \vspace{-6cm}
     \hspace{1cm}
 \begin{tabular}{l}
    \swatch{darkgoldenrod}~$\ell^+\ell^-\gamma$ \cite{ATLAS:2019gey} \\
      \swatch{orange}~$\ell^+\ell^-$+jet \cite{CMS:2019raw,ATLAS:2017nei,ATLAS:2022nrp}
    \end{tabular}
     \end{minipage} 
    \begin{minipage}{.9\linewidth}  
   
  \end{minipage} 
  \includegraphics[width=0.45\linewidth]{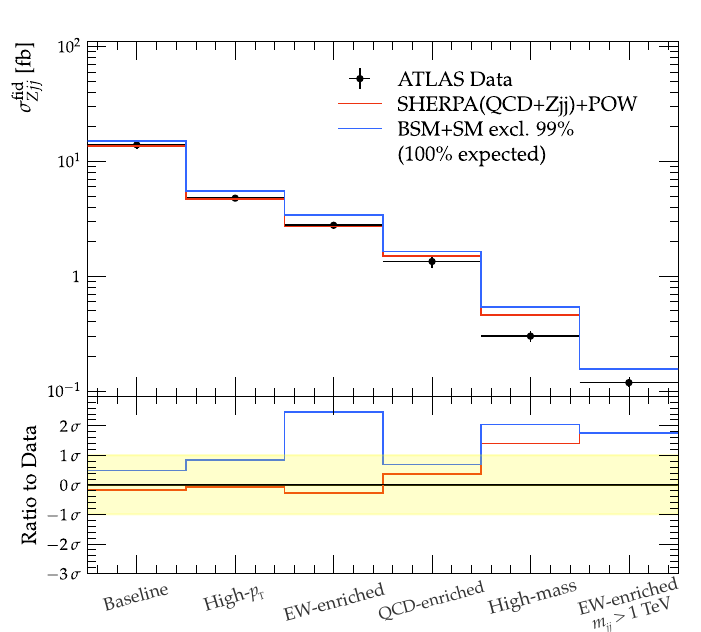}   
  \includegraphics[width=0.45\linewidth]{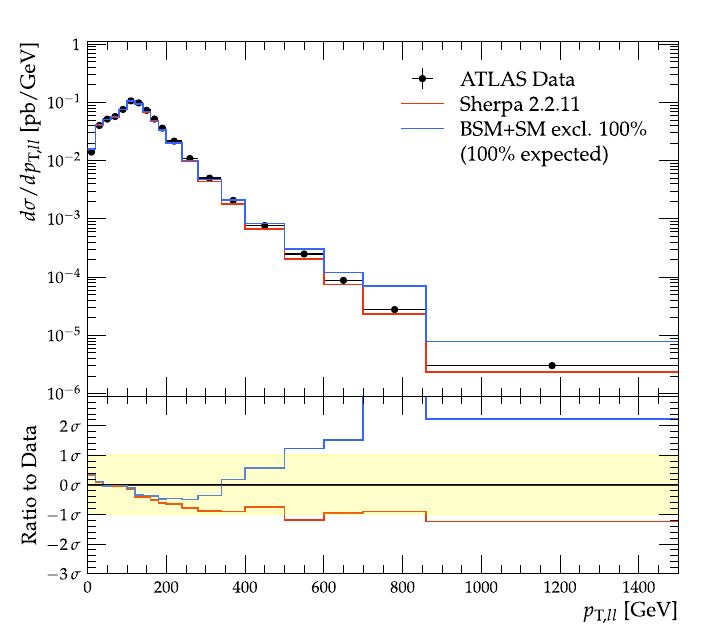}   
   \caption{(Top) Exclusion contours as a function of the cross-section times branching fraction for the production of an photo-phobic ALP with a $Z$ boson as a function of the $c\tau$, assuming an ALP mass of 40~GeV. The solid (dotted) lines represent the observed (expected) exclusions at 95\% CL. The dashed line represents the observed 68\% CL exclusion. The red dotted lines represent the 95\% CL expected exclusion extrapolated to the HL-LHC. The area above the curves is excluded. The colours of the cells represent the dominant analysis pool for each point in the grid. (Bottom left) The dominant exclusion histogram for a particular excluded point, where the cross-section is around 0.2~pb and $c\tau=0.1$~mm. The exclusion comes from an ATLAS measurement of the cross-section for electroweak production of dijets in association with a $Z$ boson~\cite{ATLAS:2017nei}, specifically, Table 3 of that paper. (Bottom right) The dominant exclusion histogram for a particular excluded point, where the cross-section is around 1~pb and $c\tau=100$~m. The exclusion comes from an ATLAS measurement $Z$ + high transverse momentum jets~\cite{ATLAS:2022nrp}, specifically, Figure 5a of that paper. The black markers represent the observed data, with the red line showing the SM predictions. The blue line shows the SM prediction summed with the new physics prediction, which in this case results in a significant disagreement with the observation.}
    \label{fig:ALP-results}
\end{figure}

For the \textsc{Contur} study, events were generated using  \textsc{MadGraph5\_aMC@NLO} v3.4.2 and \textsc{Pythia} v8 for hadronisation using a UFO file corresponding to the model presented in Ref.~\cite{Brivio:2017ije}. The event generation is limited to cases where the $Z$ decays into electrons or muons, to be consistent with the ATLAS analyses.  The grid is composed of 10 points spaced logarithmically in $\sigma \times B$ between 0.1~mm and 10~m and 10 points spaced logarithmically in $c\tau$ between 0.01~mm and 1000~m, with 10000 events generated at each point.
The exclusions obtained for this model using \textsc{Contur} are shown in Figure~\ref{fig:ALP-results} for the case where the ALP mass is 40 GeV. The exclusion is dominated by dilepton plus jet analyses everywhere in the plane, as would be expected since the final state under study consists of two light leptons and two gluons. The most sensitive analyses are the $Z$+jets measurements from ATLAS~\cite{ATLAS:2017nei} (the same one already giving exclusion for the HS case in Section~\ref{sec:HS}, but this time in the electroweak-enriched bin) and an equivalent measurement by CMS~\cite{CMS:2019raw}. In this plot, constraints can also be established in the high-lifetime regime due to a clear effect of the recoil of the ALP in the $Z$ transverse momentum spectrum measured in Ref.~\cite{ATLAS:2022nrp}.

\begin{figure}[t!]
    \centering
    \includegraphics[width=0.48\linewidth]{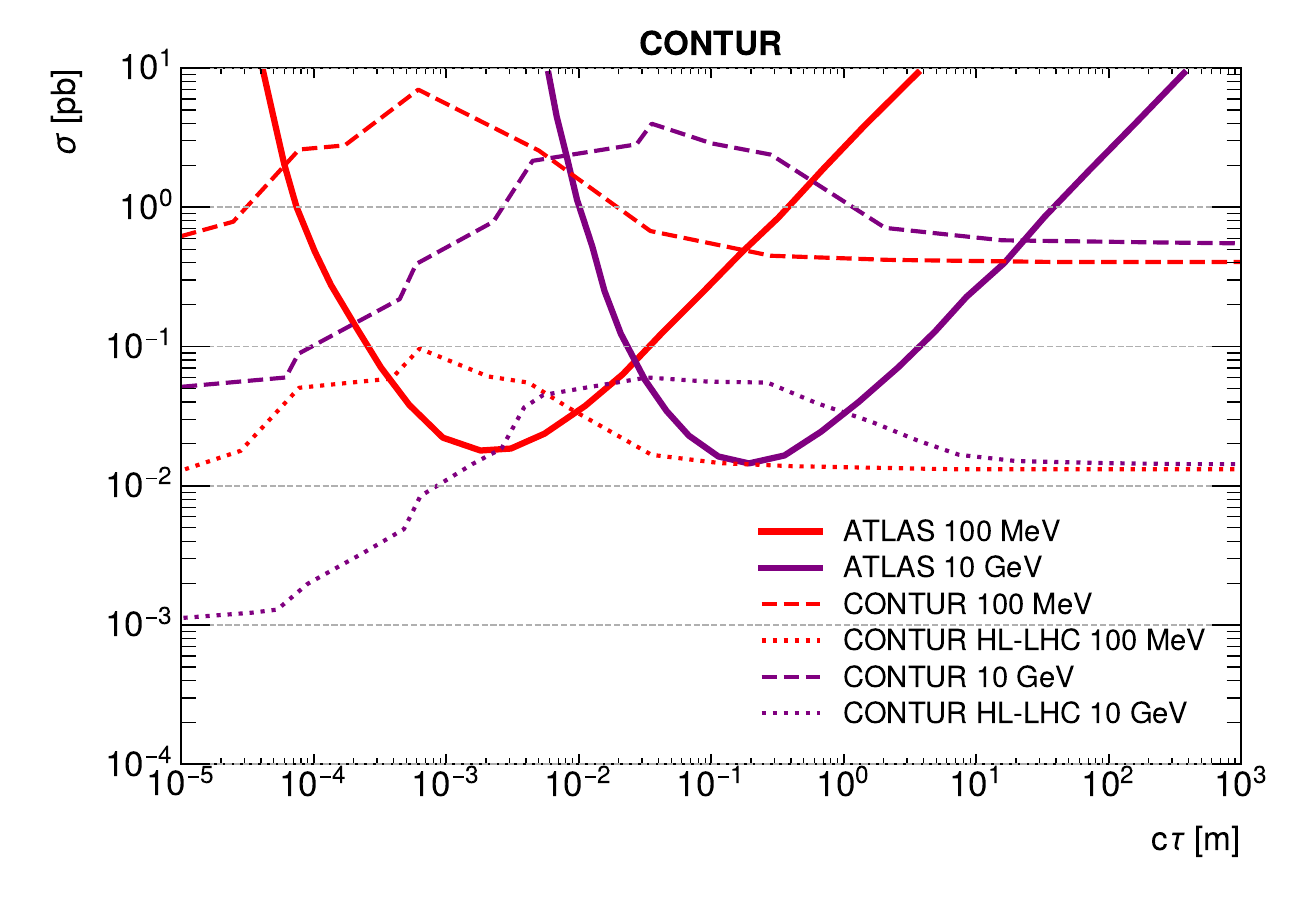}
   \includegraphics[width=0.48\linewidth]{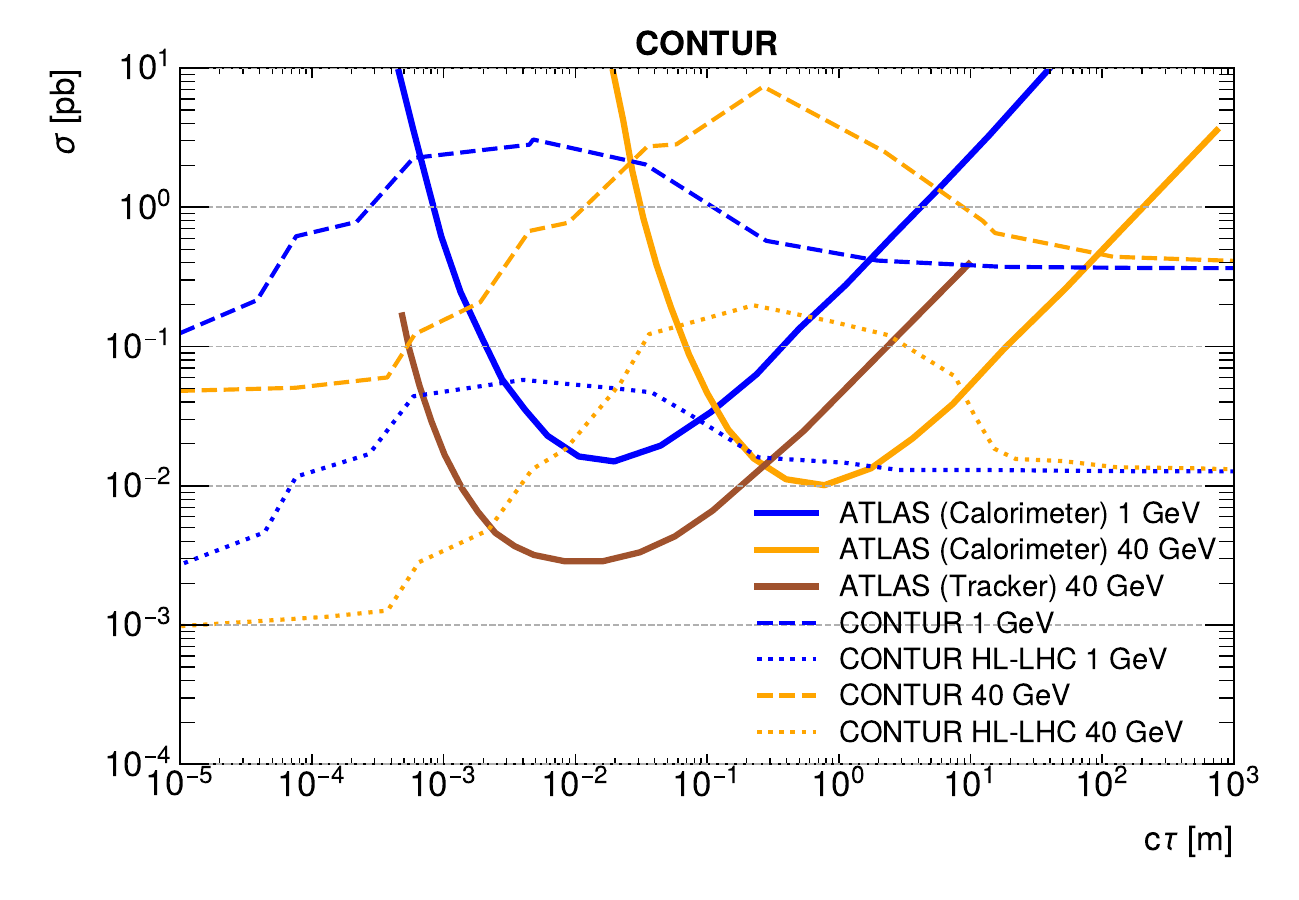}
    \caption{ Comparison of \textsc{Contur} results for a photo-phobic ALP produced in association with a $Z$ boson, extrapolated \textsc{Contur} results for HL-LHC, and dedicated search results from ATLAS exploiting the calorimeter~\cite{ATLAS:2024ocv} and tracker~\cite{ATLAS:2024qoo}.}
    \label{fig:ALP-results-summary}
\end{figure}

A comparison of the \textsc{Contur} results with the best-available constraints from ATLAS are shown in Figure~\ref{fig:ALP-results-summary}. They show that the constraints extracted from the measurements are of the same order of magnitude as the direct searches, and ruling out the sub-millimeter part of the spectrum, as well as the region above hundreds of meters, in a way that is complementary and competitive with the direct search programme. The plots also show the expected reach of measurements would increase the sensitivity by more than an order of magnitude over the course of the HL-LHC.

\newpage

\section{Conclusions}\label{sec:conclusions}

In this work, a method to constrain long-lived particle models using the fraction of particles that decay early was shown. It exploits the \textsc{Contur} method to derive constraints from the bank of LHC measurements, but filtering out particles or events which would not be reconstructible. The method was used to determine new constraints on four popular long-lived particle models of various types. The considered models are varied in the nature of the LLPs: charged or neutral, singly-produced or pair-produced, leptonically or hadronically decaying.
In each case, new parts of the parameter space can be excluded. Projections of how the constraints could evolve into the HL-LHC era are also presented, which sometimes can match or even over-take direct searches in some parts of the parameter space. These findings may be relevant when it comes to effort allocation within the ATLAS and CMS collaborations in the sense that, at least in some instances, measurements might offer comparable sensitivities to a variety of New Physics scenarios as dedicated searches for prompt objects, especially for shorter lifetimes, and in some instances, very long lifetimes. Accounting for these results could liberate human resources in order to better probe other parts of parameter space.
Moreover, the HL-LHC projections show that large parts of the so far unexplored parameter space could become accessible: in other words, that new long-lived particle discoveries could be awaiting us with the onset of Run 4 of the LHC. 

\section*{Acknowledgements}

We wish to thank Jackson Burzynski for sharing his expertise on track and vertex reconstruction, and Jon Butterworth for invaluable sanity checks. Simon Jeannot received funding from the Université Clermont Auvergne Graduate Track for Mathematics and Physics to pursue this research.
%%%%%%%%%%%%%%%%%%%%%%%%%%%%%%%%%%%%%%%%%%%%%%%%%%%%%%%%%%%%%%%%%%%%%%%%%%%%%%%%%%%
%%%%%%%%%%%%%%%%%%%%%%%%%%%%%%%%%%%%%%%%%%%%%%%%%%%%%%%%%%%%%%%%%%%%%%%%%%%%%%%%%%%
%%%%%%%%%%%%%%%%%%%%%%%%%%%%%%%%%%%%%%%%%%%%%%%%%%%%%%%%%%%%%%%%%%%%%%%%%%%%%%%%%%%

\bibliographystyle{JHEP}
\bibliography{references}

\section*{Appendix}

\begin{figure}
    \centering
       \subfloat[FIMP]{\includegraphics[width=0.45\linewidth]{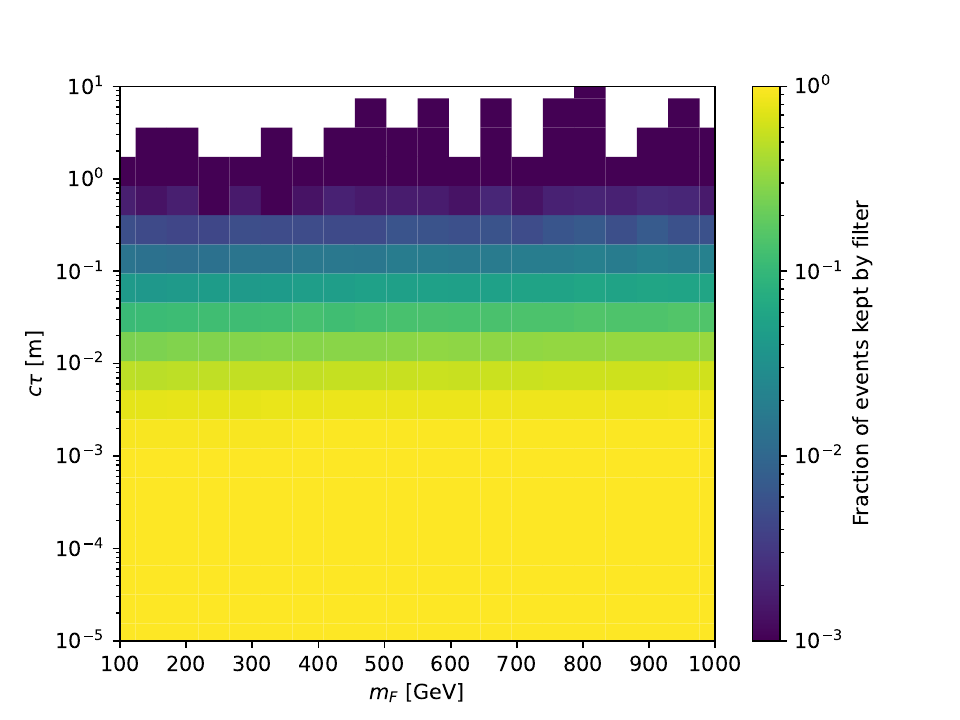}}\\
    \subfloat[HS $m_\Phi = 1000$ GeV, $m_S = 475$ GeV]{\includegraphics[width=0.45\linewidth]{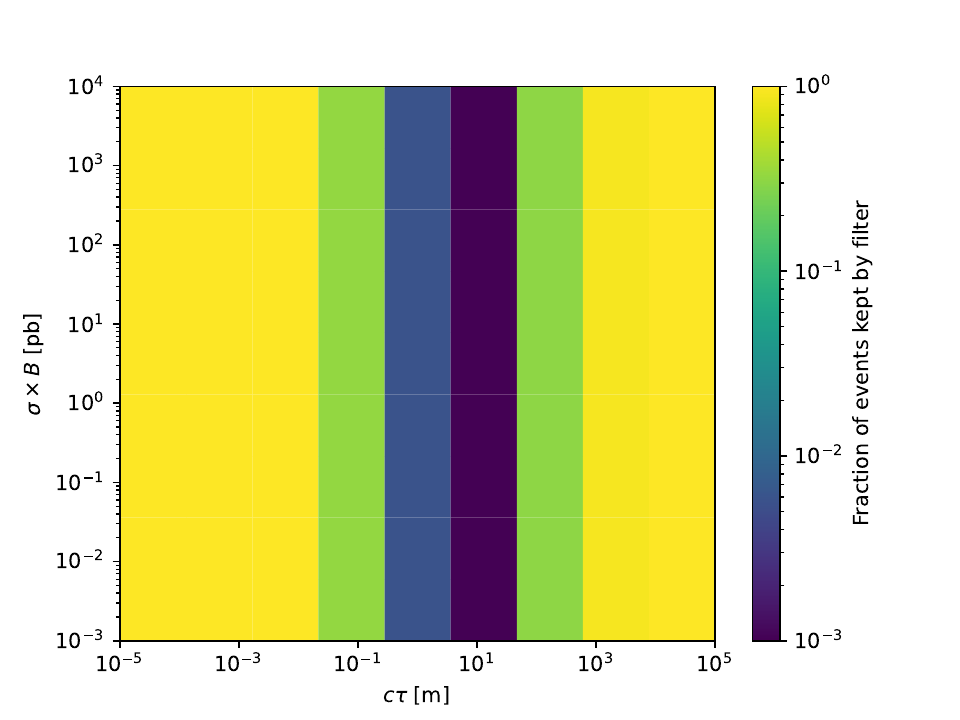}}
    \subfloat[HS $m_\Phi = 1000$ GeV, $m_S = 150$ GeV]{\includegraphics[width=0.45\linewidth]{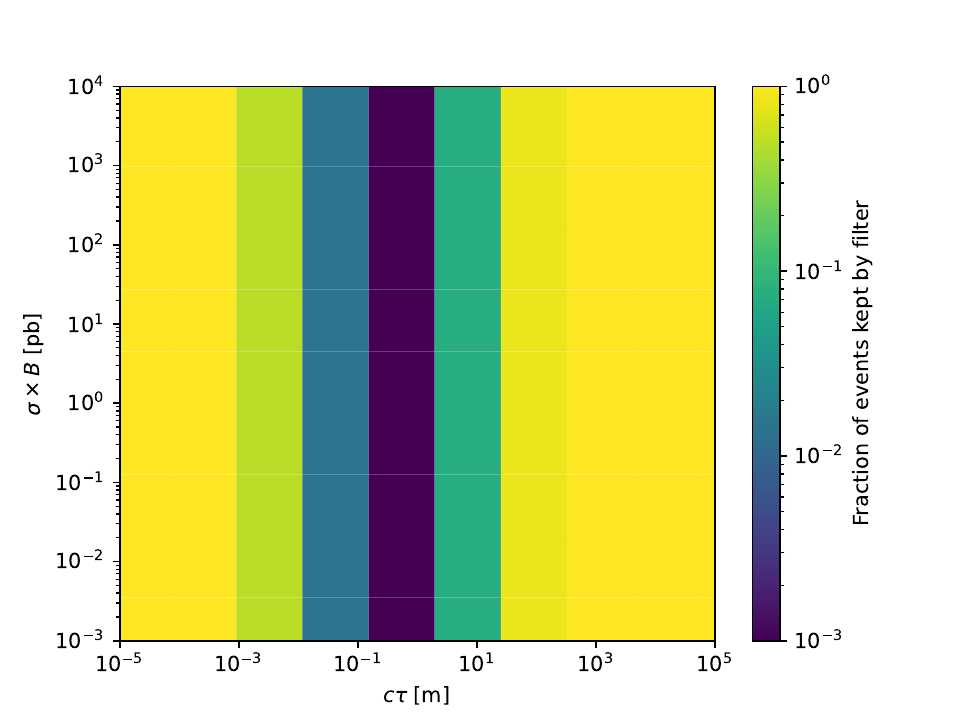}}\\
  \subfloat[Dark photon $m_\Phi = 600$ GeV, $m_{Z_d} = 400$ GeV]{\includegraphics[width=0.45\linewidth]{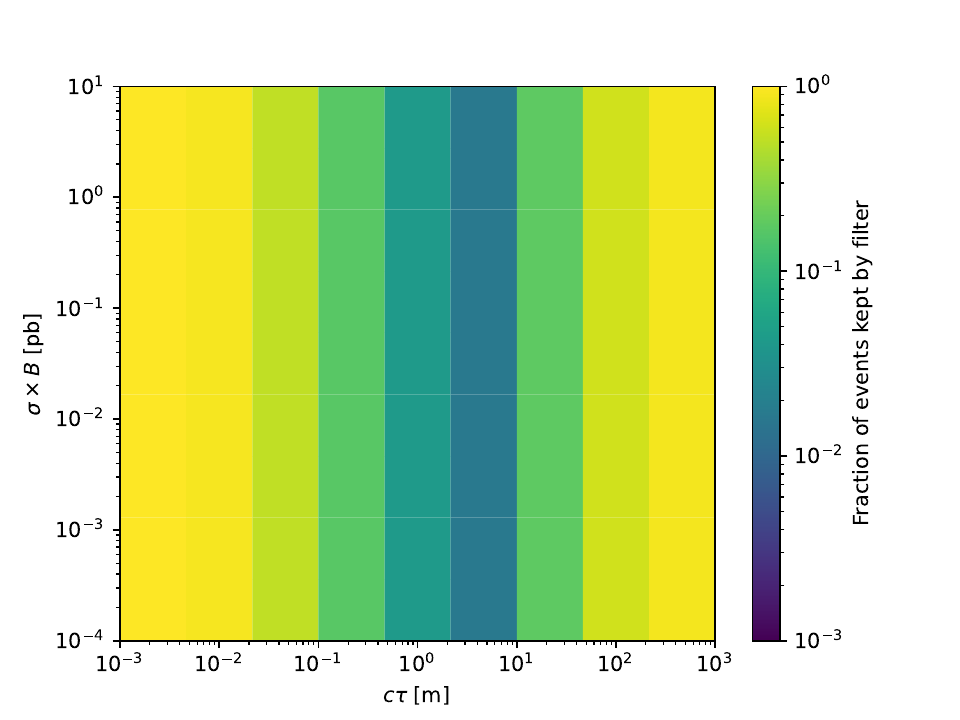}}
    \subfloat[Axion-like particle $m_\textrm{ALP} = 10$ GeV]{\includegraphics[width=0.45\linewidth]{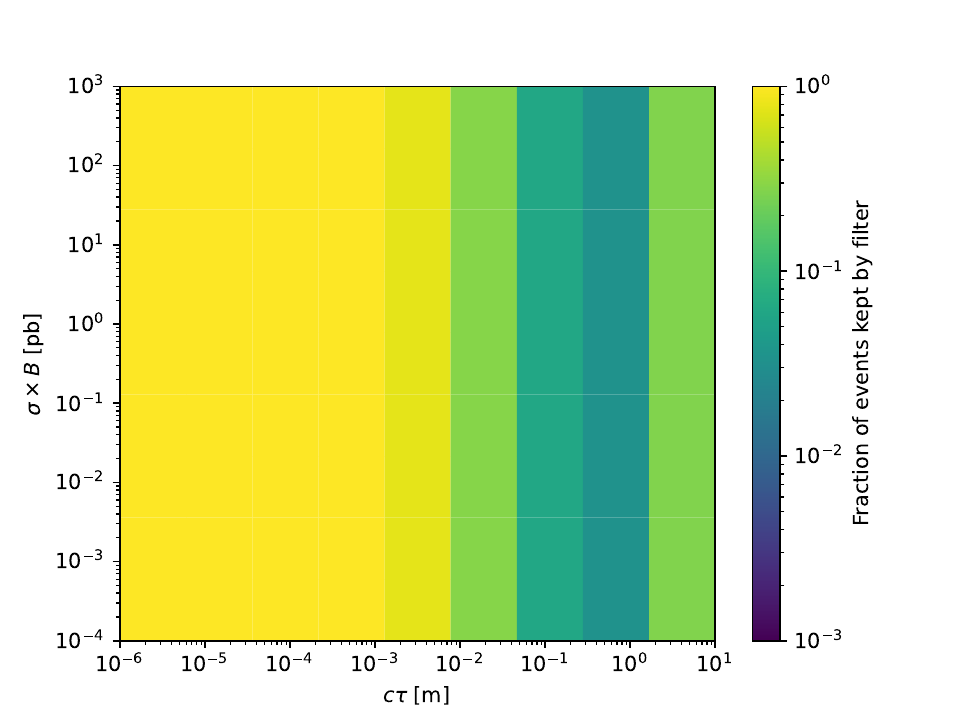}}
    \caption{The fraction of events kept by the event filter described in the main body of the text, for example parameter points for each model that was considered in this paper. For low values of c$\tau$, almost all events are kept. As $c\tau$ increases, increasing numbers of particles decay after the 5~mm threshold in transverse decay position, and are deemed un-reconstructible. In that case, the entire event is discarded. This is because ATLAS and CMS measurements may treat such events in data as instrumental noise. At high lifetimes, the events can be salvaged if the neutral LLP decay took place after the physical bounds of the detector. In those cases, the neutral LLPs are removed, and hence are treated as missing transverse energy. In the models where two LLPs are produced (a, b, c), it can be seen that the drop-off in events kept by the filter is steeper than for models where only one LLP is produced (d, e). White cells represent values of 0.}
    \label{fig:enter-label}
\end{figure}

\end{document}